\documentclass[5p,preprint,sort&compress]{elsarticle}
 
\usepackage[utf8]{inputenc}
\usepackage{siunitx}
\usepackage{graphicx}
\usepackage{amsmath,mathtools, cuted}
\usepackage{subcaption}
\usepackage[table,dvipsnames]{xcolor}
\usepackage{booktabs}
\usepackage{multirow}
\usepackage{bm}
\usepackage{textcomp}
\usepackage{float}
\usepackage{gensymb}
\usepackage{chemformula}
\usepackage{amsopn}
\usepackage{tabularx}
\usepackage{enumitem}
\usepackage{times}
\usepackage{lscape}
\usepackage{threeparttable}
\usepackage{tabularx}
\usepackage{caption}
\usepackage{pifont}
\usepackage{longtable}
\usepackage{threeparttablex}
\usepackage{listings}
\usepackage{hyperref}
\usepackage{cleveref}
\journal{Acta Materialia}

\crefname{appendix}{}{}

\hypersetup{
    hidelinks,
    colorlinks=false,
    linkcolor=black,
    citecolor=black,
    urlcolor=black
}

\begin{document}

\begin{frontmatter}

\title{Machine Learning Potentials for Alloys: A Detailed Workflow to Predict Phase Diagrams and Benchmark Accuracy}

\author[1]{Siya Zhu}
\ead{siyazhu@tamu.edu}

\author[1]{Do\u{g}uhan Sar\i{}t\"urk}
\ead{sariturk@tamu.edu}

\author[1,2,3]{Raymundo Arr\'oyave}
\ead{raymundo.arroyave@tamu.edu}

\affiliation[1]{organization={Department of Materials Science and Engineering, Texas A\&M University},
                city={College Station},
                postcode={77843}, 
                state={TX},
                country={USA}}

\affiliation[2]{organization={J. Mike Walker '66 Department of Mechanical Engineering, Texas A\&M University},
                city={College Station},
                postcode={77843}, 
                state={TX},
                country={USA}}

\affiliation[3]{organization={Wm Michael Barnes '64 Department of Industrial and Systems Engineering, Texas A\&M University},
                city={College Station},
                postcode={77843}, 
                state={TX},
                country={USA}}

\begin{abstract}
High-entropy alloys (HEAs) have attracted increasing attention due to their unique structural and functional properties. In the study of HEAs, thermodynamic properties and phase stability play a crucial role, making phase diagram calculations significantly important. However, phase diagram calculations with conventional CALPHAD assessments based on experimental or ab-initio data can be expensive. With the emergence of machine-learning interatomic potentials (MLIPs), we have developed a program named PhaseForge, which integrates MLIPs into the Alloy Theoretic Automated Toolkit (ATAT) framework using our MLIP calculation library, MaterialsFramework, to enable efficient exploration of alloy phase diagrams. Moreover, our workflow can also serve as a benchmarking tool for evaluating the quality of different MLIPs.
\end{abstract}


\end{frontmatter}

\section{Introduction}
High‐entropy alloys (HEAs) have attracted intense research interest because their vast compositional space offers unprecedented combinations of strength, ductility, and thermal stability \cite{miracle2017critical,senkov2010refractory}.  A primary requirement for alloy discovery is phase-stability prediction---knowing which phases are thermodynamically favored as functions of composition, temperature, and pressure.  Phase-diagram computations are therefore central to modern alloy design.  Although high-throughput, first-principles\allowbreak–\allowbreak based phase-diagram workflows have grown rapidly in recent years \cite{li2020high}, accurate and efficient prediction of multicomponent phase stability remains a cornerstone challenge in materials science.

Classical CALPHAD assessments \cite{lukas2007computational} provide formal, reproducible, and rigorous routes to thermodynamic databases for binary and ternary systems, with higher-order behavior inferred by extrapolation. Yet, \emph{fewer than $\sim$3{,}500 ternary phase diagrams have been experimentally determined out of roughly $1.3\times10^{5}$ possible ternaries}, meaning that only about $3\%$ of all ternary systems are even partially characterized \cite{villars1995handbook}. However, the weeks to months of expert parameter optimization they require limit the scalability and predictive power in unexplored chemical spaces, particularly those relevant to HEAs and other compositionally complex materials.  To overcome these limitations, researchers are increasingly using (1) density functional theory on special quasirandom structures (SQS) to approximate random atomic configurations \cite{zunger1990special,van2017software,samanta2024software,zhu2021computational}; (2) large-scale molecular dynamics (MD) simulations to capture temperature-dependent thermodynamic properties \cite{broughton1987phase,bai2023short}; and (3) cluster expansion Hamiltonians fitted to \emph{ab-initio} data to model configurational contributions to alloy energetics \cite{zhu2023probing,nataraj2021systematic}.  While such workflows are capable of facilitating the exploration of thermodynamic spaces that have yet to be explored experimentally, their computational cost still scales combinatorially with element count, underscoring the need for faster yet equally reliable surrogates for multicomponent phase-diagram prediction.

Recent advances in machine learning interatomic potentials (MLIPs) offer a promising alternative to accelerate phase diagram calculations by bridging quantum-mechanical accuracy with the efficiency required for large-scale thermodynamic modeling~\cite{rosenbrock2021machine}. In our previous work~\cite{zhu2025accelerating}, we introduced a benchmarking framework that demonstrated the potential of MLIPs to predict formation energies and phase stability in alloy systems with high fidelity. Although this study highlighted the feasibility of MLIP-based phase diagram computation, the methodology remained limited in scope and flexibility. In this work, we present an expanded and fully automated computational workflow that integrates MLIPs with our \textit{MaterialsFramework} library into the phase diagram calculation pipeline with the Alloy-Theoretic Automated Toolkit (ATAT)~\cite{van2002alloy,van2017software}. Our newly developed code, named \textit{PhaseForge}, introduces several key features, including support for multiple MLIP frameworks, automated structure sampling, vibrational free energy estimation, MD calculations of liquid phase, and compatibility with CALPHAD-style database generation. These capabilities enable high-throughput phase diagram predictions across a wide range of alloy systems, including those lacking prior thermodynamic assessments.

We validate our workflow through two examples---the \ch{Ni-Re} binary system and the \ch{Co-Cr-Fe-Ni-V} quinary system. The \ch{Ni-Re} system contains FCC, HCP and liquid phases, and two intermetallic compound phases with multi-sublattices are predicted. The results of several MLIPs are compared with \emph{ab-initio} calculations and experimental reports, and show how our framework can be used to benchmark the MLIPs from a thermodynamics perspective. For the \ch{Co-Cr-Fe-Ni-V} HEA system, we include all the binary and ternary sub-systems, and demonstrate the effectiveness in capturing phase stability trends in both stable and metastable regions of the phase diagram. 

Our approach highlights the growing contribution of machine learning to materials science---enabling data-driven alloy design and accelerating the exploration of previously inaccessible materials spaces. Conversely, constructing phase diagrams with various MLIPs using our workflow offers materials science a way to give back---by providing a practical and application-oriented framework for evaluating the effectiveness of MLIPs.

\section{Methods}
\subsection{Compound energy formalism}
Phase diagrams are constructed with the compound energy formalism (CEF) \cite{hillert2001compound} and exported as standard \texttt{TDB} files that downstream CALPHAD software can read without modification. In the ATAT workflow, we first generate SQSs for each disordered phase and evaluate their total energies with MLIPs; ordered phases are handled with fully relaxed stoichiometric cells. Every mixing or formation energy is referenced to the Gibbs free energies of the corresponding unary reference phases (e.g., FCC Ni, HCP Co, BCC V) provided by the SGTE database \cite{dinsdale1991sgte}.

Preserving these SGTE reference states is necessary as even a small shift would mis-rank the relative stabilities of competing phases, rendering the resulting \texttt{TDB} file CALPHAD-incompatible and undermining subsequent extrapolation to higher-order systems. By anchoring MLIP energies to the SGTE scale, our workflow guarantees that all derived interaction parameters remain thermodynamically consistent and immediately usable in conventional CALPHAD assessments and databases.

For the basic model, we assume that the Gibbs free energy contains an ideal configurational entropy of mixing with a potentially non-ideal enthalpy of mixing~\cite{van2017software}:

\begin{equation} \label{eq:nonIdealGibbs}
\begin{aligned}
      G^\beta\left(y,T\right) & = \left( E_\text{ML}^\beta \left( y \right) - \sum_i x_i E_\text{ML}^{\alpha_i}\right)\\ & + \sum_i x_i G^{\alpha_i}_\text{SGTE} \left( T\right) - TS_\text{id}\left(y,T\right)
\end{aligned}
\end{equation}

\noindent where \(E\) and \(G\) denote, respectively, the per-atom internal energy and Gibbs free energy. We include only the \emph{ideal} configurational entropy of mixing, \(S_{\text{id}}\). The symbols \(\alpha_i\) represent the unary reference phases for element \(i\) (e.g., BCC, FCC, HCP), while \(\beta\) designates the multicomponent phase under consideration. The vector \(y\) collects the site fractions of each constituent in phase \(\beta\); \(x_i\) is the overall atomic fraction of element \(i\) in the alloy. Pressure–volume (\(PV\)) work is neglected because all phases treated here are condensed and experience only negligible volume changes within the temperature range of interest.

For improved accuracy, one can add phonon contributions to the modeling for the end members of each ordered phase; in other words, we have phonon contributions that are configuration independent. The free energy is given by:

\begin{equation} \label{eq:phononGibbs}
\begin{aligned}
    G^\beta\left(y,T\right) &= \left( E_\text{ML}^\beta \left( y \right) - \sum_j w_j E_\text{ML}^\beta \left(j\right)\right)\\ 
    &+ \sum_j w^j \left( G_\text{ML}^\beta \left( y^j,T\right) - \sum_i x_i^j G_\text{ML}^{\alpha_i}\left(T\right)\right) \\
    &+ \sum_i x_i G^{\alpha_i}_\text{SGTE} \left( T\right) - TS_\text{id}\left(y,T\right)
\end{aligned}
\end{equation}
\noindent with $y^j$ the vector of site fractions of an end member $j$, and $w^j$ denotes the weights of each end member in $y$ -- $\sum_j w^jy^j=y$. The $G_\text{ML}$ are the Gibbs free energy calculated with machine learning potentials, with energy and phonon contributions.

For maximum accuracy, configuration–dependent phonon contributions can also be included. In that case, the Gibbs free energy is written as

\begin{equation} \label{eq:entropyGibbs}
\begin{aligned}
      G^\beta\left(y,T\right) & = \left( G_\text{ML,nc}^\beta \left( y \right) - \sum_i x_i G_\text{ML,nc}^{\alpha_i}\right)\\ & + \sum_i x_i G^{\alpha_i}_\text{SGTE} \left( T\right) - TS_\text{id}\left(y,T\right)
\end{aligned}
\end{equation} 
where the subscript ``nc'' indicates free energies with no configurational entropy contributions.

\subsection{Molecular dynamics for the liquid phase}
Reliable phase–diagram predictions require an accurate description of the liquid state, because the liquid sets the upper limit of thermodynamic stability at extreme operating temperatures and governs many alloy‐processing routes (e.g.\ casting, welding, additive manufacturing). To obtain these properties, we perform MD simulations to evaluate the average free energy of the liquid phase, using the ASE framework~\cite{larsen2017atomic} with Nosé–Hoover ensembles~\cite{melchionna1993hoover}. Although temperature-dependent liquid free energies of \emph{pure} elements are available from the SGTE database, MD calculations on liquid‐alloy SQS are still required to obtain the \emph{mixing} free energies. These alloy MD data place the liquid phase on the same thermodynamic footing as the solid phases, eliminating systematic bias in subsequent solidus and liquidus predictions.

For pure elements, we set the temperature $T_0 = T_m + \Delta T$, with $T_m$ the melting point and $\Delta T=\SI{50}{\kelvin}$ by default. For alloy systems, the temperature is set as linear-combination of melting points of compositions:
 
\begin{equation}
    T_0 = \sum_i x_i T_m^{\alpha_i} + \Delta T
\end{equation}

The liquid SQS configurations is pre-generated in ATAT at various compositions. As the MD simulation with NPT ensemble may converge slowly under the ASE framework by our experience, we offer two options in our code to obtain the MD energy.

An option is the ternary search in the ASE framework. To obtain the energy at equilibrium, it is necessary to optimize the volume at which the total energy reaches its minimum. The total energy is expected to decrease monotonically with increasing volume until it reaches a minimum, after which it increases monotonically. We define a scaling factor $c=l/l_0$, where $l$ and $l_0$ are the box length of the liquid structure in MD simulation and SQS, respectively. Within a range of scaling factor $\left[c_L, c_R\right]$ containing the equilibrium one (typically 0.7 to 1.0 from our experience), for each iteration, we pick $c_1 = 2c_L/3+c_R/3$ and $c_2=c_L/3+2c_R/3$, performing MD simulation with scaling factor $c_1$ and $c_2$, and comparing the total energies. If $E(c_1)>E(c_2)$, we search for the equilibrium structure within $[c_1,c_R]$ for the next iteration; otherwise, we search within $[c_L,c_2]$. After approximately six iterations, the search range of the scaling factor can be narrowed to within 0.02, with the total energy difference reduced to less than \SI{30}{\milli\eV/atom} based on our calculations. The whole process takes about 2 to 3 hours for SQS with 32 atoms on a single CPU core, depending on the MLIP used in the calculation.

Another option we implemented in our framework is to apply the MLIP to LAMMPS~\cite{plimpton1995fast,thompson2022lammps}. For each structure, we generate a $2\times2\times2$ supercell of the SQS in ATAT and run MD with the NPT ensemble at $P=0$ for \SI{30}{\pico\second}, following with an NVT ensemble for another \SI{30}{\pico\second}. The final energy is derived from the average of NVT total energies in each step.

\subsection{Phonon and mechanical instabilities}
For phonon calculations, we apply the Born von Karman spring model by fitting the reaction forces from supercells with imposed atomic displacements~\cite{van2002automating}, which is applied in the \texttt{fitfc} code in ATAT~\cite{van2009multicomponent}.

Mechanical instabilities—manifested as negative elastic constants or imaginary phonon modes—are common in solid systems. By contrast, CALPHAD assessments assume that every phase possesses a single, composition-continuous Gibbs-energy surface across the whole alloy space. A common remedy is to \emph{extrapolate} the Gibbs energy from the mechanic ally stable region into the unstable composition range, yet extrapolations taken along different paths need not coincide and can even fall \emph{below} the energy of the true stable phase. In the \ch{Co–Nb–V} ternary, for example, the BCC Gibbs energy of pure Co obtained by extrapolating the \ch{Nb–Co} binary differs from that obtained via the \ch{V–Co} binary, occasionally yielding an unrealistically low value for BCC~Co.

To eliminate such ambiguities, we adopt the \textit{ inflection detection} (ID) procedure implemented in ATAT~\cite{van2018reconciling}. ID locates the composition where the Gibbs energy curvature changes sign, the mechanical stability limit, and uses that state as a consistent thermodynamic reference, thereby avoiding ad hoc extrapolations. The ID option is disabled by default but can be activated with the \texttt{-id} flag (see \Cref{sec:ID}). We also provide pre-computed ID-corrected unary data for 31 common metallic elements in BCC, FCC, and HCP structures (\Cref{sec:pureDB}), which users may import directly into their CALPHAD workflows.

\subsection{Database for pure elements with \emph{ab-initio} calculations} \label{sec:pureDB}
The phonon computations and the inflection detection method require very accurate calculations of forces and stresses. Thus far, most MLIPs have shown good agreement with energies, but perform poorly in reproducing forces and stresses compared to \emph{ab-initio} results. Therefore, based on \emph{ab-initio} calculations, we construct a database for several common metallic elements in BCC, FCC, and HCP lattices, containing energies and vibrational entropies, including those obtained via inflection-detection for mechanically unstable phases. The \emph{ab-initio} calculations are performed with Vienna Ab-inito Simulation Package (VASP)~\cite{kresse1993ab,kresse1994ab,kresse1996efficiency,kresse1996efficient}, using the Perdew-Burke-Ernzerhof (PBE) exchange and correlation functional at level of the generalized gradient approximation (GGA)~\cite{perdew1996generalized}. The cutoff energy is set as 1.3 times the ENMAX in the pseudopotential files in the projector-augmented wave (PAW) method~\cite{blochl1994projector}. K-points per reciprocal atom is set as 8000 for all calculations. The data can be used to replace for $E^{\alpha_i}_\text{ML}$ in \Cref{eq:nonIdealGibbs}, $G^{\alpha_i}_\text{ML}\left( T \right)$ in \Cref{eq:phononGibbs}, or $G^{\alpha_i}_\text{ML,nc}\left( T \right)$ in \Cref{eq:entropyGibbs}.

In addition, for some thermodynamically stable phases with phonon instabilities under harmonic approximation, the ``dynamical stabilization'' , such as hopping between local minima, guarantees the stability at a macroscopic level. One prime example would be BCC $\beta$ titanium, which is stable above \SI{880}{\degreeCelsius}. We recommend the recent Piecewise Polynomial Potential Partitioning (P$^4$) method~\cite{kadkhodaei2017free,kadkhodaei2020software} for free energy calculations. The calculations should be performed separately and manually re-visit the energy file. A database with energies and vibrational entropies of 31 common metallic elements is constructed and illustrated in \Cref{sec:appPureAbinitio}. The mechanical instability information of the phases is adopted from \cite{van2018reconciling}, and the energies and vibrational entropies are compared with data from \cite{van2018reconciling,chen5236216computational}.

\subsection{Other modeling considerations}
As provided in ATAT, short-range order can be accounted for by adding an extra free-energy term \(G_{\text{SRO}}\) calculated with the Cluster Variation Method (CVM)~\cite{kikuchi1951theory}. In our workflow this correction is applied \emph{only} to single-sublattice phases, where it offsets the over-stabilisation of a fully disordered solid solution. For multi-sublattice phases the long-range ordering is treated explicitly and an additional SRO term would spuriously lower the energy~\cite{van2017software}. The SRO correction is therefore \textbf{off}  by default and can be enabled with the \texttt{-sro} flag.

By default, the unary end-member references are the thermodynamic data in the SGTE database. One can exclude the data from the SGTE database of particular phases by creating a file \texttt{exfromsgte.in}. For example, for BCC Al which is mechanically unstable, we may add \texttt{SGTE\_BCC\_A2\_AL} to the \texttt{exfromsgte.in} to exclude the SGTE data, and the $G^\text{BCC}\left( T \right)$ would be denoted as 
\begin{equation}
    G^\text{BCC}\left( T \right) = G_\text{calc}^\text{BCC}\left (T \right) - G_\text{calc}^\text{FCC}\left (T \right) +G_\text{SGTE}^\text{FCC}\left (T \right)
\end{equation}
instead of $G_\text{SGTE}^\text{BCC}\left( T \right)$, where the ``calc'' denotes the free energies calculated with MLIPs, or obtained from our database calculated with \emph{ab-initio}. This would be useful for phases that are mechanically unstable and where inflection detection is implemented, as it may lead to a discrepancy between SGTE data and calculations.

\section{Usage}

The overall workflow of phase diagram calculation using ATAT and \textit{PhaseForge} would be: 
\begin{enumerate}[label=(\roman*)]
    \item (Optional) Incorporate the MLIP within \textit{PhaseForge} (or use the MLIPs pre-integrated);
    \item (Optional) Construct SQS for phases/levels not included in ATAT database;
    \item Choose the desired elements and phases;
    \item Calculating the relaxed energies (and vibrational free energies if needed) of each structure using MLIP;
    \item (Optional) Perform molecular dynamics for the liquid phase using MLIP;
    \item Fit a CALPHAD model (\texttt{TDB} file) for the system.
\end{enumerate}
For the following, replace the text within brackets [ ] with appropriate user input, excluding the brackets themselves. In \Cref{sec:CLI} we provide a simple in-line command for quick phase diagram calculation. In \Crefrange{sec:MLIPsInCLI}{sec:fitting} we introduce the detailed process. 

\subsection{A quick inline command for phase diagram calculation} \label{sec:CLI}
In \textit{PhaseForge}, we provide a quick inline command:
\begin{lstlisting}
sqscal -e [Element1, Element2, ...] -l [Lattice1, Lattice2] -lv [Level] -mlip [MLIP] -model [MLIP model version] [-vib] [-sro]
\end{lstlisting}

where [Element1,Element2,…] is the list of elements symbols in the system (the element should be included in the MLIP), the [Lattice1, Lattice2] are the standard CALPHAD crystal structure names (e.g. FCC\_A1, BCC\_A2), [Level] is the fineness of composition grid, [MLIP] is the machine-learning interatomic potential used in the calculation, and [MLIP model version] indicates the specific model version. [-vib] option indicates the vibrational contribution is calculated (for endmembers), and the [-sro] option indicates the short-range order correction is applied. In this command, a ternary-search is applied for liquid with a temperature offset of \SI{50}{K}, and binary interactions to the level 1 for each phase are included by default. One can modify the source code or follow the step by step workflow outlined in \Crefrange{sec:MLIPsInCLI}{sec:fitting} if modifications to the default parameters are needed. 

\subsection{Incorporate MLIP within \textit{MaterialsFramework}} \label{sec:MLIPsInCLI}

We achieved the integration of MLIP into our workflow using \textit{MaterialsFramework}. The framework’s modular infrastructure enabled us to incorporate a selected MLIP directly into the computational pipeline with ease. Although we used a single MLIP model for all example calculations in this study, the design of \textit{MaterialsFramework} makes it straightforward to substitute different MLIPs without altering the core workflow. This flexibility is a powerful feature, allowing future users to efficiently benchmark and compare MLIPs or tailor potential selection to specific alloy systems and property prediction goals.

\begin{table}[t]
\begin{threeparttable}
\centering
\caption{The list of MLIPs currently implemented in \textit{MaterialsFramework}, along with the default versions specified for each model. The table includes version identifiers as used in the framework and references to the original publications or official documentation describing each model.}
\label{tab:MLIP}
\begin{tabularx}{\columnwidth}{lXl}
\toprule
\textbf{MLIP Name} & \textbf{Version} & \textbf{Reference} \\ \midrule
Alignn & v12.2.2024\_dft\_3d\_307k & \cite{choudhary_atomistic_2021,choudhary_unified_2023} \\ \midrule
AlphaNet\tnote{$\dagger$} & AlphaNet-oma-v1 & \cite{yin_alphanet_2025} \\ \midrule
CHGNet & 0.3.0 & \cite{deng_chgnet_2023} \\ \midrule
DeePMD\tnote{$\dagger$} & DPA3-v2-OpenLAM & \cite{wang_deepmd-kit_2018,zeng_deepmd-kit_2023,zeng_deepmd-kit_2025} \\ \midrule
EqV2\tnote{$\dagger$} & eqV2-L-OA & \cite{barroso-luque_open_2024} \\ \midrule
eSEN\tnote{$\dagger$} & eSEN-30M-OA & \cite{fu_learning_2025} \\ \midrule
GPTFF\tnote{$\dagger$} & gptff\_v2 & \cite{xie_gptff_2024} \\ \midrule
Grace & GRACE-2L-OMAT & \cite{bochkarev_graph_2024} \\ \midrule
HIENet\tnote{$\dagger$} & HIENet-0 & \cite{yan_materials_2025} \\ \midrule
M3GNet & MP-2021.2.8-PES & \cite{chen_universal_2022} \\ \midrule
MACE & mace\_mpa\_0 & \cite{batatia_foundation_2024} \\ \midrule
MatterSim & mattersim-v1.0.0-5m & \cite{yang_mattersim_2024} \\ \midrule
NewtonNet & t1x & \cite{haghighatlari_newtonnet_2022} \\ \midrule
ORB & orb-v3 & \cite{neumann2024orbfastscalableneural,rhodes2025orbv3atomisticsimulationscale} \\ \midrule
PetMad\tnote{$\dagger$} & v1.0.1 & \cite{mazitov_pet-mad_2025} \\ \midrule
PosEGNN\tnote{$\dagger$} & pos-egnn.v1-6M &  \cite{ibm_pos_egnn} \\ \midrule
SevenNet & SevenNet-MF-ompa & \cite{park_scalable_2024,kim_data-efficient_2025} \\ \bottomrule
\end{tabularx}
\begin{tablenotes}
  \item [$\dagger$] These models require manual download of their respective checkpoint or parameter files due to licensing or distribution restrictions. Please refer to the official repositories for instructions on obtaining and configuring these files.
\end{tablenotes}
\end{threeparttable}
\end{table}

\textit{MaterialsFramework} supports a broad suite of state-of-the-art MLIPs—including GRACE, Eqv2, ORB, eSEN, and others\allowbreak–\allowbreak that have demonstrated strong performance in predicting structural, thermodynamic, and mechanical properties across diverse alloy systems. \Cref{tab:MLIP} lists the currently available models in \textit{MaterialsFramework} along with their respective default versions. A more comprehensive list of all available versions for each model is provided in \Cref{sec:appMLIPversions}. These models are integrated via pre-configured calculators that seamlessly interface with our SQS-based pipeline, enabling efficient energy and force evaluations without the computational cost of \emph{ab-initio} methods for structure relaxation, vibrational entropy calculations, and MD simulations. This modular design ensures that MLIP benchmarking and substitution can be conducted with minimal overhead, encouraging comparative analysis.

\subsection{Construct Special Quasirandom Structures} \label{sec:mkSQS}
The ATAT database includes over 30 common crystal structures and their SQS (including the initial structure of liquid phase for the molecular dynamics). To generate SQS for structures not included in ATAT, or generate SQS with different level or size, we use the \texttt{mcsqs} module in ATAT~\cite{van2009multicomponent}. First we create a folder \texttt{atat/data/sqsdb/[Lattice\_name]} for the structure we need. Two input files are required: a \texttt{rndstr.skel} file that defines the structure, and a \texttt{sqsgen.in} file that specifies the levels and compositions on each sublattice. An example for D0$_{19}$ structure is in \Cref{sec:appD019}.

 With the two input files, we use command
\begin{lstlisting}
sqs2tdb -mk
\end{lstlisting}
to create a series of folders for each SQS, with a file named \texttt{rndstr.in} in each folder. In each folder, using the \texttt{corrdump} command in ATAT:
\begin{lstlisting}
corrdump -l=rndstr.in -ro -noe -nop -clus -2=... -3=...
\end{lstlisting}
we can generate the clusters within the given range of pairs, triplets, etc. in a file named \texttt{clusters.out}. Then we use
\begin{lstlisting}
mcsqs -n=[Number of atoms]
\end{lstlisting}
to start the Monte Carlo process searching for the SQS. The number of atoms in the supercell SQS should be a multiple of atomic sites in the primitive cell (we suggest that an SQS with 30 to 50 atoms per cell would be suitable). The \texttt{mcsqs} code stops if it finds a perfect match in \texttt{bestsqs.out} for all the correlations. Sometimes the perfect match may not exist, as we have too many clusters or the supercell is not large enough. In this case, we can manually stop the code by \texttt{touch stopsqs} and the \texttt{bestsqs.out} contains the best matched structure found currently. When the best SQS's are constructed in each folder under \texttt{atat/data/sqsdb/[Lattice\_name]}, we are prepared for the next step.

\subsection{Structures and elements selection} \label{eq:selection}
With MLIP and SQS prepared, we can use the \texttt{sqs2tdb} command in ATAT for calculation. The first step is copying the SQS from the ATAT database (or generated in \Cref{sec:mkSQS}):
\begin{lstlisting}
sqs2tdb -cp -sp=[Element1, Element2, ...] -l=[Lattice_name] -lv=[Level]
\end{lstlisting}
It creates a folder with the lattice name and a file named \\ \texttt{species.in} in the folder. After confirming the elements, retype the same command to create a series of folders named \texttt{sqsdb\_lev=[Level]\_[sublattice\_a]\_[Element\_A]=\allowbreak[Concentration of A in sublattice a]...} for each SQS. In each folder, we have a structure file \texttt{str.out} in ATAT format. 

\subsection{Relaxed energies calculations using MLIP} \label{sec:relaxedEnergies}
To relax the structure and calculate the free energy of each SQS, first we need to generate an empty dummy \texttt{vasp.wrap} file in the top folder which is required by ATAT. Use the command:
\begin{lstlisting}
runstruct_vasp -nr
\end{lstlisting}
we can get the standard VASP input files \texttt{INCAR}, \texttt{POSCAR}, \texttt{POTCAR}, and \texttt{KPOINTS}. It may give error information for the missing pseudopotentials for VASP and we can just neglect it as we only need the \texttt{POSCAR} file for MLIP calculations. To relax the structure, we use 
\begin{lstlisting}
MLIPrelax -mlip=[MLIP] -model=[MLIP model version]
\end{lstlisting}
It generates a Python script \texttt{MLIPrelax.py} and launches it for the structural relaxation and calculations of energy, forces, and stresses. By using the \texttt{extract\_MLIP} command after it, the relaxed structure is in \texttt{CONTCAR} and converted to the ATAT format in \texttt{str\_relax.out}. The energy, stress and forces are in \texttt{energy}, \texttt{stress.out} and \texttt{force.out}, respectively. 

To make it more convenient, use command \texttt{runstruct\_mlip -mlip=[MLIP name] -model=[MLIP model version]} , which does the following:
\begin{lstlisting}
runstruct_vasp -nr
MLIPrelax -mlip=[MLIP] -model=[MLIP model version]
extract_mlip
\end{lstlisting}

\subsection{Inflection Detection for mechanically unstable phases} \label{sec:ID}
For mechanically unstable phases, we follow "robust relax" process with "inflection-detection" method applied in ATAT. We provide a command \texttt{robustrelax\_mlip} like the \\
\texttt{robustrelax\_vasp} in ATAT. A recommanded usage of the command is:
\begin{lstlisting}
robustrelax_mlip -id -c=0.05 -mlip=[MLIP] -model=[MLIP model version]
\end{lstlisting}
which means relaxing the structure and using inflection-detection method with a criteria relaxation magnitude cutoff as 0.05. We can use it to replace the command \texttt{runstruct\_mlip} in \Cref{sec:relaxedEnergies}. 

The inflection-detection process requires not only the accurate energies, but also the forces and stresses of the calculations. It might fail or find the wrong inflection point with poor accuracy in forces and stresses with the MLIP. To overcome this, we provide a pure metallic elements database in \Cref{sec:appPureAbinitio}, with some common metallic elements in BCC, FCC and HCP. Inflection detections are done with VASP, with the energies and vibrational entropies pre-calculated and saved in the database.

\subsection{Vibrational Entropy Calculations}\label{sec:vibrationalEntropy}
We use the \texttt{fitfc} module in ATAT to calculate the vibrational entropy. After generating an empty dummy \texttt{vaspf.wrap} file, one can do:
\begin{lstlisting}
foreachfile endmem pwd \; fitfc -si=str_relax.out -ernn=4 -ns=1 -nrr
foreachfile -d 3 wait \; runstruct_vasp -lu -w vaspf.wrap -nr
foreachfile -d 3 wait \; MLIPcalc -mlip=[MLIP] -model=[MLIP model version]
foreachfile endmem pwd \; fitfc -si=str_relax.out -f -frnn=1.5
foreachfile endmem pwd \; robustrelax_vasp -vib
\end{lstlisting}
for the calculation of the vibrational entropy for each end member at high temperature limit in \texttt{svib\_ht}, in units of Bolzmann constants.

Sometimes we might have negative frequencies and the \texttt{fitfc} code aborts with the error information:
\begin{lstlisting}
Unstable modes found. Aborting...
\end{lstlisting}
If you see the warning:
\begin{lstlisting}
  Warning: p... is an unstable mode.  
\end{lstlisting}
then the structure is certainly unstable. Otherwise it could be an artifact of the fitting procedure. Several ways to deal with the issue:
\begin{itemize}
    \item Use \texttt{fitfc -fu} to find and check the unstable mode. It generates a file \texttt{unstable.out } under folder \texttt{vol\_0}. The unstable modes are outputed in form of: \texttt{u [index] [nb\_atom] [kpoint] [branch] [frequency] [displacements...]}. If the file only contains \\ \texttt{nb\_atoms} as \texttt{too\_large}, you need to increase the \texttt{-mau} option. Otherwise, you can pick one of the unstable modes and do \texttt{fitfc -gu=[unstable mode index]} to generate the unstable mode. Run the single-point calculation with \texttt{MLIPcalc} in the generated subdirectory (named \texttt{vol\_0/p\_uns\_*} and rerun \texttt{fitfc -f -fr=...}. Repeat the process until the error message disappear or a truly unstable mode is found.
    
    \item Use a larger supercell for the calculation;
    \item Decrease the \texttt{-frnn};
    \item If the structure is mechanically unstable, follow the steps in \Cref{sec:ID} for inflection detection;
    \item If one believe the structure is mechanically stable, use the option \texttt{-fn} in \texttt{fitfc -si=str\_relax.out -f -frnn=2 -fn} to force the \texttt{fitfc} code to calculate the vibrational entropy.
\end{itemize}
The vibrational contribution calculations require accurate forces from the calculations. When the MLIP you use are poor in forces and stresses calculations, we recommand using the pure metallic elements database introduced in \Cref{sec:ID} and \Cref{sec:appPureAbinitio} if available.

\subsection{Molecular Dynamics calculations for liquid phase} \label{sec:mdLiquid}
For the liquid phase, we use molecular dynamics to calculate the free energies at different concentrations. Use the the command:
\begin{lstlisting}
sqs2tdb -cp -sp=[Element1, Element2, ...] -l=LIQUID -lv=[Level]
\end{lstlisting}
to generate initial liquid "SQS" for MD calculations. With an empty dummy \texttt{vasp.wrap} file, in each folder, use the command:
\begin{lstlisting}
runstruct_vasp -nr
MLIPliquid -mlip=[MLIP] -model=[MLIP model version] -dT=[Temperature offset, default=50] [-LAMMPS]
\end{lstlisting}

It first calculates the temperature for MD, which is the linear combination of the melting points of all the elements in the structure, plus the offset temperature passing by the argument, which guarantees the final temperature is above melting point and we get a liquid structure in MD. The code performs a ternary search for the equilibrium volume and calculates the energy, or alternatively computes it using LAMMPS with the \texttt{[-LAMMPS]} option. A sample LAMMPS input script for MD is provided in \Cref{sec:lammpsscript}. 

\subsection{Fitting into a CALPHAD model} \label{sec:fitting}
With all the energies of SQS's and vibrational entropies (if needed) calculated, we can use ATAT to fit the data into a CALPHAD model. For each phase, we use the command:
\begin{lstlisting}
sqs2tdb -fit
\end{lstlisting}
It will generate a file named \texttt{terms.in}, containing the levels of interactions included in the \texttt{TDB} file. It is in the format:
\begin{lstlisting}
    order,level: order,level:...
 ...
\end{lstlisting}
 The \texttt{order} is 1 for the linear combination term, 2 for binary interaction, 3 for ternary interaction, etc., and the \texttt{level} indicates the level of polynomials included for that order of interaction. Starting from \texttt{order=1}, the \texttt{order,level} pairs should be included in the \texttt{terms.in}, with the same order in each line, and separated by colons for each sublattice. For
 \begin{itemize}
     \item \texttt{order=1}, we can only have \texttt{level=0};
     \item \texttt{order=2}, the binary interaction contribution to the free energy is $\sum_{l=0}^LG_{AB;l}X_AX_B\left(X_B-X_A\right)^l$, where the $L$ is the \texttt{level} of polynomial;
     \item \texttt{order=3},\texttt{level=0} indicates a single term in form of $G_{ABC}X_AX_BX_C$, while  higher levels contain extra terms $G_{ABC;1}X_AX_BX_C\left(V_AX_A+V_BX_B+V_CX_C\right)$;
     \item \texttt{order>3}, we only have \texttt{level=0}.
 \end{itemize}

 With the \texttt{terms.in} file set properly, retype the command
\begin{lstlisting}
sqs2tdb -fit [-sro]
\end{lstlisting}
to perform the fitting process, with the \texttt{-sro} optional, which includes the short-range-order contribution to the model. There would be an output file named \texttt{lattice\_name.tdb} in the folder. After the fitting of each lattice is completed, we can combine the results of all the \texttt{TDB} files, together with SGTE elemental data, with a command in the base folder:
\begin{lstlisting}
sqs2tdb -tdb [-oc]
\end{lstlisting}
It generates a \texttt{TDB} file named \texttt{[ELEMENT\_ELEMENT\_...].tdb}, which can be imported into thermodynamic packages using Calphad modeling, such as Thermo-Calc~\cite{sundman1985thermo,ANDERSSON2002273}, Pandat~\cite{chen2002pandat}, FactSage~\cite{BALE20161}, OpenCalphad~\cite{sundman2015opencalphad}, etc. to plot phase diagrams. The \texttt{[oc]} option is the Open Calphad option for a more portable \texttt{TDB} file.

\section{Examples}

In this work, we consider the \ch{Ni-Re} system and the \ch{Co-Cr\allowbreak-Fe-Ni-V} system as examples to demonstrate our method. The \ch{Ni–Re} binary system contains potential intermetallic compounds with multiple sub-lattices, and several MLIPs are applied to compare with VASP results, illustrating how our method can benchmark MLIPs from a materials science perspective. The \ch{Co–Cr–Fe–Ni–V} HEA system demonstrates how our method can handle multi-element systems and highlights its efficiency compared to traditional \emph{ab-initio} calculations.

\subsection{\ch{Ni-Re} Binary System}
\begin{figure*}[htbp]
    \centering
    \includegraphics[width=\textwidth]{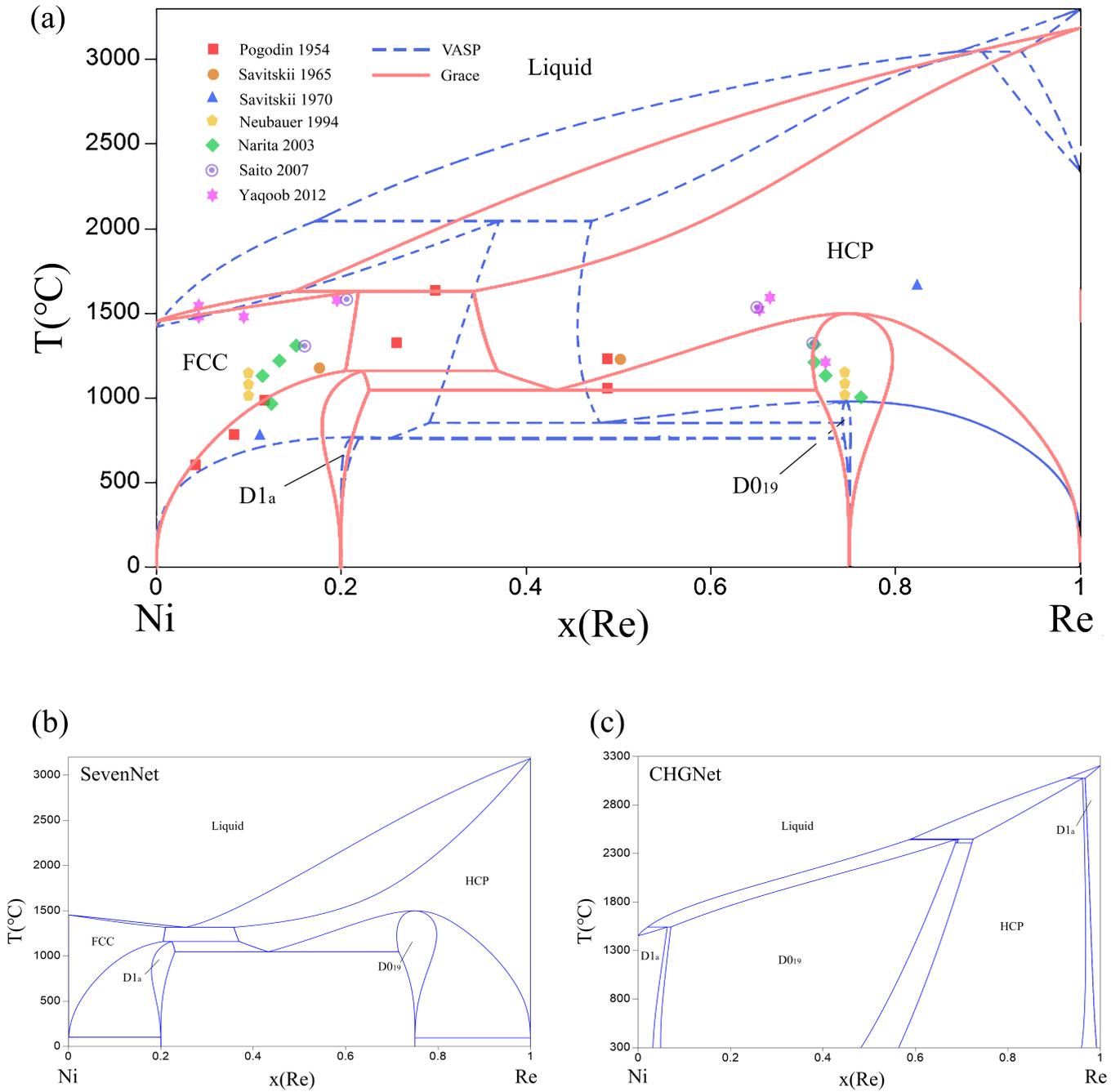}
    \caption{(a) Phase diagram of \ch{Ni-Re} system with FCC\_A1, HCP\_A3, D0$_{19}$, D1$_a$, and liquid phases, calculated with VASP and Grace; (b) Phase diagram calculated with SevenNet; (c) Phase diagram calculated with CHGNet.}
    \label{fig:NiRe-GraceVasp}
\end{figure*}

\ch{Ni-Re} alloys are widely used in aerospace engineering owing to their robust strength under high-temperature conditions. Levy \textit{et al.}~predicted several novel intermetallic compounds in Re-based alloys, including two potential new phases in the \ch{Ni-Re} system - the D0$_{19}$ NiRe$_3$ phase and the D1$_a$ Ni$_4$Re phase. In our previous work\cite{zhu2021computational}, we used ATAT in combination with VASP to investigate potential intermetallic compound phases. A phase diagram including the FCC\_A1, HCP\_A3, D0$_{19}$, D1$_a$, and liquid phases was constructed. In this work, we apply the new MLIP-based framework to reproduce the phase diagram with improved efficiency. The detailed workflow is presented below:
\begin{itemize}
    \item Construct the SQS of D0$_{19}$ and D1$_a$ phases;
    \item Generate \ch{Ni-Re} SQS of various phases and compositions with ATAT;
    \item Optimize the structures and calculate energies at \SI{0}{K} using MLIP;
    \item Perform MD simulations on liquid phase of different compositions (with ternary search);
    \item Fit all the energies with CALPHAD modeling using ATAT;
    \item Construct the phase diagram with Pandat.
\end{itemize}
A \texttt{terms.in} 
\begin{lstlisting}
1,0
2,0
\end{lstlisting}
is applied for the FCC\_A1, HCP\_A3 and liquid phases to include the binary interactions to the level 0, while a \texttt{terms.in} 
\begin{lstlisting}
1,0:1,0
2,0:1,0
\end{lstlisting}
is applied to include only the binary interactions on a single sub-lattice to the level 0, for the D0$_{19}$ and the D1$_a$ phases. Vibrational contributions are not included in our calculations, as some MLIPs may exhibit insufficient precision in forces and stresses calculations. In \Cref{fig:NiRe-GraceVasp}(a), we illustrate the phase diagram calculated with VASP (in dashed blue lines) and Grace (with Grace-2L-OMAT model, in red solid lines), with some results from experiments previously reported~\cite{yaqoob2012experimental,pogodin1954nickel,savitskii1965rhenium,savitskii1970alloys,neubauer1994diffusion,saito2007phase,okamoto2012ni,Narita2003}. The result calculated with Grace captures most of the topology successfully and shows good agreements with VASP result, in spite of more stability of the intermetallic compounds, and a lower peritectic temperature of FCC\_A1 and HCP\_A3 (\SI{2044}{\celsius} from VASP and \SI{1631}{\celsius} from Grace). In addition, the allotropic transition for pure Re at high temperature, which is an artifact of using a simple harmonic model for the phonon free energy, disappears in Grace results, since no vibrational contributions are considered. It should be noted that, although the Grace result appears to show better agreement with experimental data, this does not indicate a higher accuracy compared with VASP. MLIPs are trained on energies from \emph{ab-initio} calculations, and the apparent match in this case may result from a coincidental cancellation of errors arising from the MLIP modeling, the energy database, the ATAT workflow, and the CALPHAD modeling.

In \Cref{fig:NiRe-GraceVasp}(b) and (c), we illustrate the phase diagram calculated with SevenNet (version SevenNet-MF-ompa) and CHGNet (version 0.3.0). The SevenNet result successfully captures most of the phase diagram topology, while the equilibrium temperature of FCC-HCP-Liquid is significantly underestimated, resulting in a peritectic transformation \( 
\text{L} + \text{HCP} \rightarrow \text{FCC} \) being misrepresented as eutectic \( \text{L} \rightarrow \text{FCC} + \text{HCP} \). For CHGNet, the energies are calculated with large errors; therefore, the phase diagram is largely inconsistent with thermodynamic expectations. This example shows how phase diagram computations with our workflow may serve as an effective tool to assess and compare the quality of different MLIPs.

\subsection{\ch{Co-Cr-Fe-Ni-V} HEA System}

HEA systems based on the \ch{Co-Cr-Fe-Ni-V} are attracting increasing attention, and several previous studies have explored their thermodynamic properties both experimentally and computationally~\cite{choi2019thermodynamic,he2017solid,ding2017high}. However, experimental studies can be costly, and some computational approaches rely on thermodynamic databases and expensive \emph{ab-initio} calculations (compared to MLIP-based methods). In the calculation, we only consider the FCC\_A1, BCC\_A2, HCP\_A3 solid phases and the liquid phase. No intermetallic compounds are taken into account as that would mess the phase diagram with multi-elements. All the SQS's are previously generated and employed in ATAT. For all the structure relaxations and energy calculations on SQS, we use the Grace-2L-OMAT model~\cite{bochkarev2024graph} for MLIP calculations, with the force convergence criterion f\_max = 0.001. For liquid phase, we use the ternary search methods from range 0.7 to 1.0 with a target of 0.02. MD simulations of the liquid phase are performed with ASE using the Grace-2L-OMAT model, with an NVT ensemble for each structure in ternary search for 2ps. First-principles results of pure elements in all solid phases are adapted from our database and shown in \Cref{tab:abpure}. Mechanically unstable phases including FCC-Cr, BCC-Co, BCC-Ni and FCC-V have energy and vibrational entropy calculated with inflection detection method with VASP and previously added in our database. The thermodynamic data for these four phases with mechanical instability from SGTE database are excluded by creating an \texttt{exfromsgte.in} file. To fit all the thermodynamic data, we use a \texttt{terms.in} 
\begin{lstlisting}
2,1
3,0
\end{lstlisting}
to include the binary interactions to the level 1 and the ternary interactions to the level 0 for solid phases. For liquid phase, we only consider binary interactions to the level 0 (regular solution model). Short-range ordering corrections are included for all the solid phases. With the \texttt{TDB} file created with our method, we use Pandat~\cite{chen2002pandat} to plot the phase diagrams.

\begin{figure*}[htbp]
    \centering
    \includegraphics[width=\textwidth]{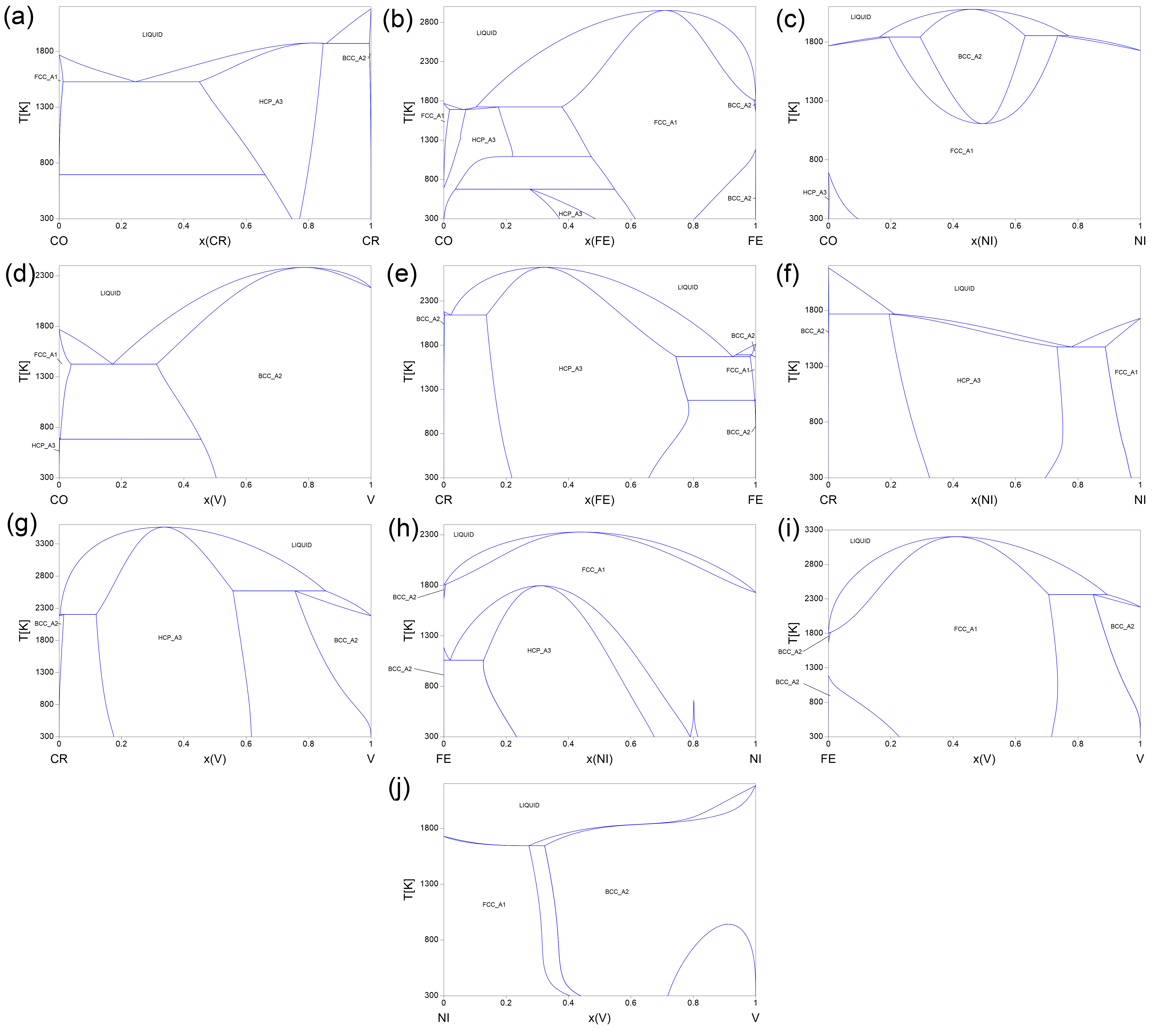}
    \caption{The 10 binary phase diagrams in \ch{Co-Cr-Fe-Ni-V} system.}
    \label{fig:binary}
\end{figure*}

\begin{figure*}[htbp]
    \centering
    \includegraphics[width=\textwidth]{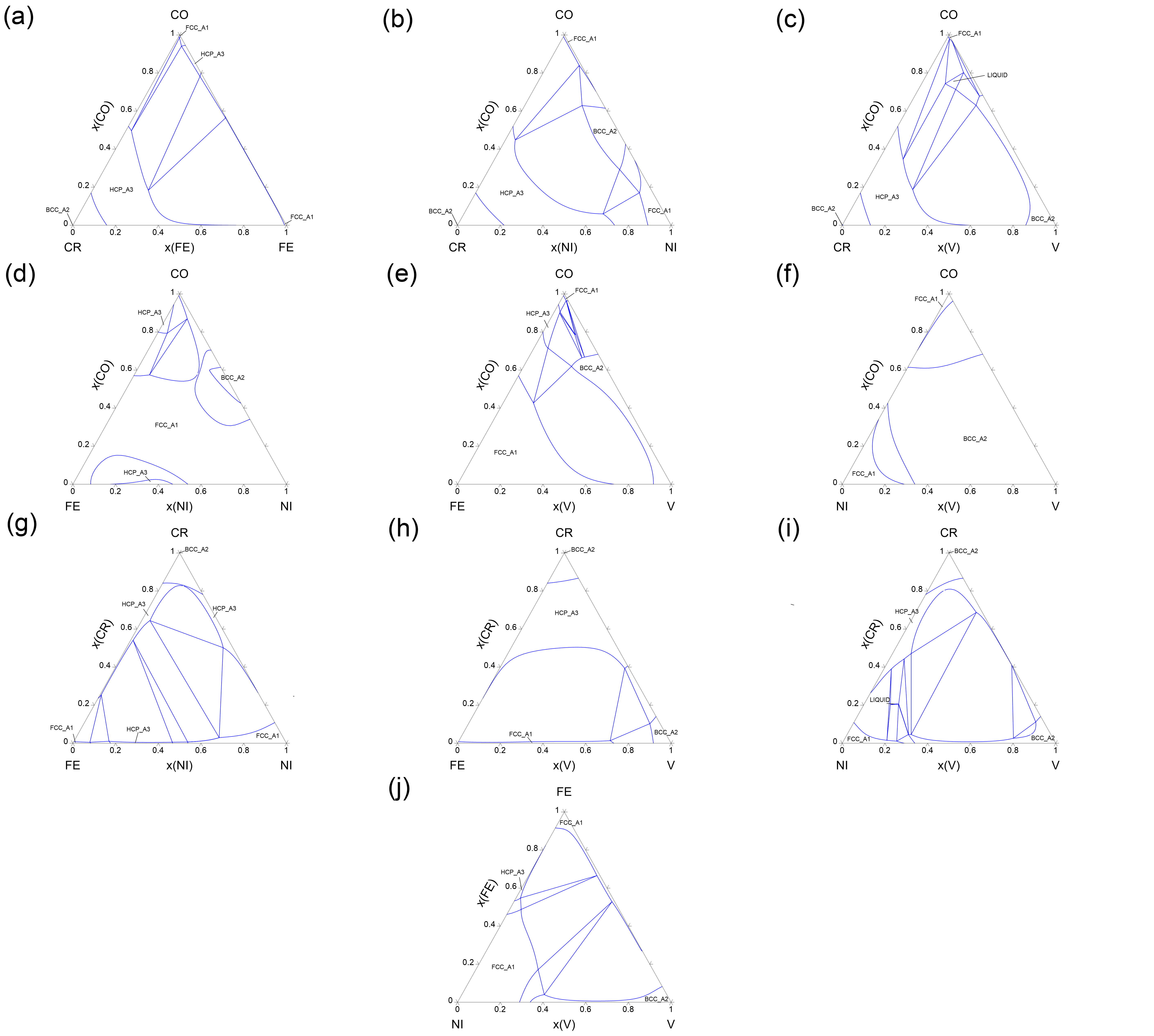}
    \caption{The 10 ternary phase diagrams in \ch{Co-Cr-Fe-Ni-V} system at \SI{1400}{\kelvin}.}
    \label{fig:ternary}
\end{figure*}

In \Cref{fig:binary}, we present all 10 binary phase diagrams. Compared with previously reported experimental and computational results~\cite{nishizawa1983co,smith1982ni,okamoto2006fe,swartzendruber1991fe,ghosh2002thermodynamic,byeong1993revision,okamoto2007co,ohnuma2002phase,okamoto2003co}, most of our predicted phase diagrams show good agreement. Despite the omission of intermetallic compounds in our analysis, the overall topology of most phase diagrams is successfully captured by our method. However, there are some issues to note. First, the HCP phase appears to be overstabilized, particularly in systems containing Cr. To verify the HCP phase energies, we take the \ch{Co–Cr} system as an example. We examined the relaxed SQS structures and their energies for all three solid phases. For \ch{Co_{0.25}Cr_{0.75}} at \SI{0}{\kelvin}, the energies are \SI{-8.8017}{\eV/atom} for HCP, \SI{-8.7986}{\eV/atom} for FCC, and \SI{-8.7949}{\eV/atom} for BCC. The energy difference is smaller than \SI{0.007}{\eV/atom}, or about \SI{675}{\joule/mol}. The small energy difference at \SI{0}{\kelvin} indicates that the instability of HCP phase at high temperature is due to the vibrational entropy. Since HCP Cr is mechanically unstable~\cite{van2018reconciling}, we use the inflection-detection method with VASP to compute its energy and vibrational entropy. For all phonon contributions, we calculate only the pure end members and approximate intermediate compositions using linear interpolation, in order to reduce computational cost and simplify the model. In other words, for SQS such as HCP \ch{Co_{0.25}Cr_{0.75}}, the energy is obtained from MLIP calculations of HCP, while the vibrational entropy is derived from a linear combination of those for HCP Co and inflection point of Cr, which may contribute to the observed discrepancies with experiments~\cite{okamoto2003co}. Besides, we do not consider intermetallic compounds like the $\sigma$ phase in \ch{Co-Cr} system, which is observed in experiments and may potentially conceal the existence of HCP. The similar issue also arises in the \ch{Co-Ni} phase diagram~\cite{nishizawa1983co}, where a single phase region of BCC wrongly appears at high temperature between x(Ni)=0.4 to x(Ni)=0.6, as both BCC Co and Ni are mechanically unstable and we used inflection detection method for the energy and vibrational contributions. Careful inflection-detection and vibrational entropy calculations on each SQS could resolve this issue, but they require significantly more computational time and demand high accuracy of the MLIP model—not only for energies, but also for forces and stresses. \Cref{fig:ternary} presents all the ternary phase diagrams generated by our workflow within the quinary system.
In \Cref{fig:ternary}, we illustrate all the 10 ternary phase diagrams at temperature of \SI{1400}{\kelvin}. 

For a reference of time cost, the relaxation and energy calculation on a SQS (from 32 atoms to 48 atoms for all the phases) takes 1 to 4 minutes calculation on a single CPU core. The ternary search of MD calculations for one structure (with 32 to 48 atoms depending on the composition) takes about 2-3 hours on a single CPU core. In total we have 70 solid SQS for each solid phase and 75 liquid structures, therefore the whole process takes less than 10 hours for only the solid phases, and takes about 200 hours for all the calculations including the liquid phase, all on a single CPU core. For contrast, relaxation and energy calculations on one SQS using VASP would take from several hours to tens of hours on a single node with 8 cores, which is thousands of the CPU time of our method.

\section{Conclusion}
In this work, we introduce our code \textit{PhaseForge} and \textit{MaterialsFramework} for exploring phase diagrams through CALPHAD modeling, integrating ATAT and MLIPs. Several novel features are implemented and introduced in detail, including energy fitting that combines data from \emph{ab-initio} calculations, MLIP predictions, and the SGTE database, as well as liquid phase fitting using molecular dynamics with MLIPs. In addition, we present the \ch{Ni-Re} system and the \ch{Co-Cr-Fe-Ni-V} HEA system as case studies.

Compared to phase diagram studies based on experiments or \emph{ab-initio} calculations, our method is significantly more efficient — typical phase diagram calculations using the ATAT–VASP workflow may require thousands of times more CPU time than our approach. On the other hand, compared to calculations based on existing thermodynamic databases, our method enables users to investigate thermodynamic properties of potential new phases or multi-element interactions that are not yet included in existing databases. Our method can be applied to high-throughput calculations of HEA phase diagrams, benefiting from its wide applicability and high efficiency. In future work, UMAP-based visualization of single-phase regions could also be applied to our computational results. The thermodynamic results can also be implemented as an initial guess or reference for active learning frameworks, and be iteratively refined with data from experiments and other calculations~\cite{kusne2020fly,ament2021autonomous,dai2020efficient}.

Nonetheless, our method represents a trade-off between computational efficiency and accuracy. Due to possible inaccuracies in the MLIP-predicted energies, stresses, and forces, the predicted phase diagram topology—particularly at high temperatures—may deviate from reality. Therefore, it is important to carefully examine the phase energies in the generated \texttt{TDB} file. 

However, the existence of inaccuracy is not always a drawback. Generating phase diagrams with various MLIPs using our framework and comparing the inaccuracy offers a new perspective for evaluating the performance of MLIPs. In materials science, it is often the energy differences, rather than the absolute energy values, that determine the phase stability and carry greater significance. Traditional evaluations by machine learning researchers typically involve comparing predicted energies, forces and stresses with reference databases to assess accuracy. In contrast, our method enables the construction of phase diagrams, allowing even an MLIP that exhibits systematic bias in absolute energies to demonstrate its usefulness—if it can still produce correct phase stability and topology. Conversely, we might also lower our assessment of MLIPs that fail to correctly predict phase diagrams. While such models may accurately reproduce absolute energy values, they can misrepresent the relative ordering of phase energies, leading to incorrect predictions of phase stability—and thus may be of limited value in materials science applications. This highlights a more application-driven criterion for evaluating MLIPs in materials science.

\section*{Code Availability}

The \textit{PhaseForge} code is publicly available at \url{https://github.com/dogusariturk/PhaseForge} (DOI: 10.5281/\allowbreak zenodo.15702127), and the \textit{MaterialsFramework} code is publicly available at \url{https://github.com/dogusariturk/MaterialsFramework} (DOI: 10.5281/zenodo.15702498).

\section*{Acknowledgements}
The authors would like to acknowledge the support of the National Science Foundation through Grant No. 2119103. We also acknowledge the support from the Army Research Office through Grant No. W911NF-22-2-0117. Calculations were carried out at Texas A\&M High-Performance Research Computing (HPRC) Facility.

\medskip
\bibliographystyle{unsrt}

\appendix
\clearpage
\onecolumn
\setcounter{table}{0}
\setcounter{page}{1}
\setcounter{figure}{0}
\setcounter{equation}{0}

\renewcommand{\thepage}{s\arabic{page}}
\renewcommand{\thetable}{S\arabic{table}}
\renewcommand{\thefigure}{S\arabic{figure}}

\section{MLIPs and versions in \textit{MaterialsFramework}} \label{sec:appMLIPversions}

\begin{ThreePartTable}
\renewcommand{\arraystretch}{1.2}
\begin{TableNotes}
  \item[$\dagger$] These models require manual download of their respective checkpoint or parameter files due to licensing or distribution restrictions. Please refer to the official repositories for instructions on obtaining and configuring these files.
\end{TableNotes}

\begin{longtable}{lcl}
\caption{Comprehensive list of MLIPs currently implemented in \textit{MaterialsFramework}, along with the versions specified for each model. The table includes version identifiers as used in the framework and references to the original publications or official documentation describing each model.} \label{tab:MLIP_SI} \\
\toprule
\textbf{MLIP Name} & \textbf{Version} & \textbf{Reference} \\
\midrule
\endfirsthead

\multicolumn{3}{c}{\tablename\ \thetable{} -- \textit{continued from previous page}} \\
\toprule
\textbf{MLIP Name} & \textbf{Version} & \textbf{Reference} \\
\midrule
\endhead

\midrule
\multicolumn{3}{r}{\textit{Continued on next page}} \\
\endfoot

\bottomrule
\insertTableNotes
\endlastfoot

Alignn & v12.2.2024\_dft\_3d\_307k & \cite{choudhary_atomistic_2021,choudhary_unified_2023} \\ \midrule
\multirow{2}{*}{AlphaNet\tnote{$\dagger$}} & AlphaNet-oma-v1           & \multirow{2}{*}{\cite{yin_alphanet_2025}}                                              \\
                                           & AlphaNet-MPtrj-v1         &                                                                                        \\ \midrule
\multirow{2}{*}{CHGNet}                    & 0.3.0                     & \multirow{2}{*}{\cite{deng_chgnet_2023}}                                               \\
                                           & 0.2.0                     &                                                                                        \\ \midrule
\multirow{2}{*}{DeePMD\tnote{$\dagger$}}   & DPA3-v2-OpenLAM           & \multirow{2}{*}{\cite{wang_deepmd-kit_2018,zeng_deepmd-kit_2023,zeng_deepmd-kit_2025}} \\
                                           & DPA3-v2-MPtrj             &                                                                                        \\ \midrule
\multirow{10}{*}{EqV2\tnote{$\dagger$}}    & eqV2-L-OAM                & \multirow{10}{*}{\cite{barroso-luque_open_2024}}                                       \\
                                           & eqV2-M-OAM                &                                                                                        \\
                                           & eqV2-S-OAM                &                                                                                        \\
                                           & eqV2-L-OMat               &                                                                                        \\
                                           & eqV2-M-OMat               &                                                                                        \\
                                           & eqV2-S-OMat               &                                                                                        \\
                                           & eqV2-L-DeNS               &                                                                                        \\
                                           & eqV2-M-DeNS               &                                                                                        \\
                                           & eqV2-S-DeNS               &                                                                                        \\
                                           & eqV2-S                    &                                                                                        \\ \midrule
\multirow{3}{*}{eSEN\tnote{$\dagger$}}     & eSEN-30M-OAM              & \multirow{3}{*}{\cite{fu_learning_2025}}                                               \\
                                           & eSEN-30M-OMat             &                                                                                        \\
                                           & eSEN-30M-MP               &                                                                                        \\ \midrule
\multirow{2}{*}{GPTFF\tnote{$\dagger$}}    & gptff\_v2                 & \multirow{2}{*}{\cite{xie_gptff_2024}}                                                 \\
                                           & gptff\_v1                 &                                                                                        \\ \midrule
\multirow{7}{*}{Grace}                     & GRACE-2L-OMAT             & \multirow{7}{*}{\cite{bochkarev_graph_2024}}                                           \\
                                           & GRACE-2L-OAM              &                                                                                        \\
                                           & GRACE-1L-OMAT             &                                                                                        \\
                                           & GRACE-1L-OAM              &                                                                                        \\
                                           & GRACE-2L-MP-r6            &                                                                                        \\
                                           & GRACE-2L-MP-r5            &                                                                                        \\
                                           & GRACE-1L-MP-r6            &                                                                                        \\ \midrule
 HIENet\tnote{$\dagger$}                   & HIENet-0                  & \cite{yan_materials_2025}                                                                        \\ \midrule
 \multirow{4}{*}{M3GNet}                   & MP-2021.2.8-PES           & \multirow{4}{*}{\cite{chen_universal_2022}}                                                      \\
                                           & MatPES-PBE-v2025.1-PES    &                                                                                                \\
                                           & MatPES-r2SCAN-v2025.1-PES &                                                                                                \\
                                           & 2021.2.8-DIRECT-PES       &                                                                                                \\ \pagebreak
 \multirow{7}{*}{MACE}                     & mace\_matpes\_0           & \multirow{7}{*}{\cite{batatia_foundation_2024}}                                                  \\
                                           & mace\_omat\_0             &                                                                                                \\
                                           & mace\_mpa\_0              &                                                                                                \\
                                           & mace\_mp\_0b3             &                                                                                                \\
                                           & mace\_mp\_0b2             &                                                                                                \\
                                           & mace\_mp\_0b              &                                                                                                \\
                                           & mace\_mp\_0a              &                                                                                                \\ \midrule
 \multirow{2}{*}{MatterSim}                & mattersim-v1.0.0-5m       & \multirow{2}{*}{\cite{yang_mattersim_2024}}                                                      \\
                                           & mattersim-v1.0.0-1m       &                                                                                                \\ \midrule
 \multirow{3}{*}{NewtonNet}                & ani1                      & \multirow{3}{*}{\cite{haghighatlari_newtonnet_2022}}                                             \\
                                           & ani1x                     &                                                                                                \\
                                           & t1x                       &                                                                                                \\ \midrule
 \multirow{13}{*}{ORB}                      & orb-v3-conservative-20-omat & \multirow{4}{*}{\cite{neumann2024orbfastscalableneural,rhodes2025orbv3atomisticsimulationscale}} \\
                                           & orb-v3-conservative-inf-omat &                                                                                                \\
                                           & orb-v3-direct-20-omat &                                                                                                \\
                                           & orb-v3-direct-inf-omat &                                                                                                \\
                                           & orb-v3-conservative-20-mpa &                                                                                                \\
                                           & orb-v3-conservative-inf-mpa &                                                                                                \\
                                           & orb-v3-direct-20-mpa &                                                                                                \\
                                           & orb-v3-direct-inf-mpa &                                                                                                \\
                                           & orb-v2 &                                                                                                \\
                                           & orb-d3-v2 &                                                                                                \\
                                           & orb-d3-sm-v2 &                                                                                                \\
                                           & orb-d3-xs-v2 &                                                                                                \\
                                           & orb-mptraj-only-v2 &                                                                                                \\ \midrule
 \multirow{3}{*}{PetMad\tnote{$\dagger$}}  & v1.1.0                    & \multirow{3}{*}{\cite{mazitov_pet-mad_2025}}                                                     \\
                                           & v1.0.1                    &                                                                                                \\
                                           & v1.0.0                    &                                                                                                \\ \midrule
 PosEGNN\tnote{$\dagger$}                  & pos-egnn.v1-6M            & \cite{ibm_pos_egnn}                                                                              \\ \midrule
 \multirow{4}{*}{SevenNet}                 & SevenNet-MF-ompa          & \multirow{4}{*}{\cite{park_scalable_2024,kim_data-efficient_2025}}                               \\
                                           & SevenNet-omat             &                                                                                                \\
                                           & SevenNet-l3i5             &                                                                                                \\
                                           & SevenNet-0                &                                                                                                \\
\end{longtable}
\end{ThreePartTable}
\clearpage

\section{Example of D0$_{19}$ structure for SQS construction} \label{sec:appD019}
The input file of the D0$_{19}$ structure \texttt{rndstr.skel} is:
\begin{lstlisting}
5.305 5.305 4.242 90 90 120   'Coordinate system
1 0 0
0 1 0
0 0 1                          'Primitive unit cell vectors
0.3333 0.6667 0.25 b           'Atom sites in the cell for sublattice 'b'
0.6667 0.3333 0.75 b
0.8333 0.6667 0.25 a           'Atom sites in the cell for sublattice 'a'
0.8333 0.1667 0.25 a
0.1667 0.8333 0.25 a
0.1667 0.3333 0.75 a
0.6667 0.8333 0.75 a
0.1667 0.8333 0.75 a
\end{lstlisting}
where a and b indicate the different sublattices. The \texttt{sqsgen.in} is:
\begin{lstlisting}
level=0 a=1         b=1
level=1 a=0.5,0.5   b=1
level=1 a=1         b=0.5,0.5
level=2 a=0.5,0.5   b=0.5,0.5
level=3 a=0.75,0.25 b=1
level=3 a=1         b=0.75,0.25
\end{lstlisting}
In this case, for level=2, we have both a and b sublattice occupied by two elements, with $x^a(A)=x^a(B) = 0.5$, $x^b(C)=x^b(D) = 0.5$, where A, B, C and D are the elements in sublattice a and b. The output file \texttt{bestsqs.out} is:
\begin{lstlisting}
5.305000 0.000000 0.000000
-2.652500 4.594265 0.000000
0.000000 0.000000 4.242000
0.000000 -1.000000 0.000000
0.000000 0.000000 -2.000000
2.000000 0.000000 0.000000
0.333300 -0.333300 -0.750000 b_A
1.333300 -0.333300 -0.750000 b_B
0.333300 -0.333300 -1.750000 b_B
1.333300 -0.333300 -1.750000 b_A
0.666700 -0.666700 -0.250000 b_B
1.666700 -0.666700 -0.250000 b_A
0.666700 -0.666700 -1.250000 b_B
1.666700 -0.666700 -1.250000 b_A
0.833300 -0.333300 -0.750000 a_B
1.833300 -0.333300 -0.750000 a_A
0.833300 -0.333300 -1.750000 a_B
1.833300 -0.333300 -1.750000 a_A
0.833300 -0.833300 -0.750000 a_A
1.833300 -0.833300 -0.750000 a_B
0.833300 -0.833300 -1.750000 a_A
1.833300 -0.833300 -1.750000 a_B
0.333300 -0.833300 -0.750000 a_A
1.333300 -0.833300 -0.750000 a_B
0.333300 -0.833300 -1.750000 a_B
1.333300 -0.833300 -1.750000 a_A
0.166700 -0.666700 -0.250000 a_B
1.166700 -0.666700 -0.250000 a_B
0.166700 -0.666700 -1.250000 a_A
1.166700 -0.666700 -1.250000 a_A
0.666700 -0.166700 -0.250000 a_A
1.666700 -0.166700 -0.250000 a_B
0.666700 -0.166700 -1.250000 a_A
1.666700 -0.166700 -1.250000 a_B
0.166700 -0.166700 -0.250000 a_B
1.166700 -0.166700 -0.250000 a_B
0.166700 -0.166700 -1.250000 a_A
1.166700 -0.166700 -1.250000 a_A
\end{lstlisting}
for the SQS of $x_A=x_B=0.5$ in both sublattices. 

\section{Ab-initio data of energy and vibrational entropy of pure metallic elements} \label{sec:appPureAbinitio}

\begin{landscape}
\begin{table}[H]
\centering
\caption{Ab-initio data of energy and vibrational entropy of pure metallic elements}
\begin{threeparttable}
\label{tab:abpure}
\begin{tabular}{c | c c c |c c c| c c c}
\toprule
\textbf{Element} & & \textbf{FCC\_A1} & & & \textbf{BCC\_A2} & & & \textbf{HCP\_A3}\\
& \textbf {Unstable} & \textbf{Energy (eV)}   & \textbf{Vibrational Entropy } & \textbf {Unstable} & \textbf{Energy (eV)}   & \textbf{Vibrational Entropy } & \textbf {Unstable} & \textbf{Energy (eV)}   & \textbf{Vibrational Entropy}  \\

\midrule

Ag & & -2.72433 & -86.1889 & & -2.61169 & -84.5878 & & -5.4422 &-172.172\\
Al & & -3.74135 & -88.2663 & \ding{51} &-3.69447 & -87.8629 & & -7.42034 & -176.257\\
Au & & -3.22782 & -85.3757 & & -3.207595 & -86.3942 & & -6.44402 & -170.539\\
Co & & -7.02531	& -88.0018 & \ding{51} & -6.9442 & -87.7826 & & -14.0868 & -176.362\\
Cr & \ding{51} & -9.32554 & -88.8351 & & -9.49173 & -88.8181 & \ding{51} & -18.5863 & -176.763\\
Cu & & -3.73145 & -87.5273 & & -3.69644 & -87.2028 & & -7.44798 & -174.943\\
Fe & & -8.08568 & -88.3182 & & -8.24337 & -88.4823 & & -16.1316 & -176.131\\
Hf & & -9.88344 & -85.6048 & & -9.7776 & -85.7277 & & -19.9156 & -172.170\\
In & & -2.55594 & -84.4606 & & -2.54688 & -84.2564 & & -5.10406 & -168.515\\
Ir & & -8.85517 & -87.0129 & \ding{51} & -8.23357 & -87.2042 & & -17.5588 & -173.746\\
Mg & & -1.49287 & -87.5432 & & -1.47669 & -87.3609 & & -3.01029& -175.256\\
Mn & & -8.90193 & -87.8455 & & -8.82125 & -87.5310 & & -17.8653 & -175.886\\
Mo & \ding{51} & -10.711 & -87.0004 & & -10.9427 & -87.9353 & \ding{51}& -21.4129 & -174.970\\
Nb & \ding{51}& -10.0079 & -87.0338 & & -10.2089 & -87.1036 & \ding{51}& -10.1419 & -173.937\\
Ni & & -5.47533 & -88.1199 & \ding{51} & -5.38133 & -87.9685 & & -10.9017 & -175.971\\
Os & & -11.0983 & -87.096 & \ding{51} & -11.0050 &-87.8551 & & -22.478 & -174.983\\
Pb & & -3.5709 & -83.8365 & \ding{51} & -3.5358 & -84.1784 & & -7.11132 & -167.472\\
Pd & & -5.21951 & -87.0289 & & -5.17659 & -86.7851 & & -10.3838 & -173.718\\
Pt & & -6.10114 & -86.4436 & \ding{51} & -6.05414 & -86.6885 & & -12.0903 & -172.131 \\
Re & & -12.2638 & -85.9419 & \ding{51}& -12.2426 & -87.3593 & & -24.7992 & -173.732\\
Rh & & -7.28059 & -87.5303 & \ding{51} & -7.06104 & -87.2504 & & -14.4864 & -175.004\\
Ru & & -9.14488 & -87.5171 & \ding{51} & -9.0875 & -87.5499 & & -18.5230 & -176.085\\
Sc & & -6.15546 & -87.1803 & \ding{51} & -6.1522 & -89.4009 & & -12.4064
 & -178.000\\
Sn & & -3.77427 & -84.2970 & & -3.76344 & -84.2074 & & -7.55217 & -168.786\\
Ta & \ding{51}& -11.6899 & -86.5449 & & -11.8579 & -86.3414 & \ding{51} & -23.3963 & -172.704\\
Ti & & -7.7081 & -87.2747 & $^\dagger$ & & & & -15.5295 & -175.495\\
V &  \ding{51} &-8.90428 & -87.8334 & & -8.94585 & -87.6379 & \ding{51}& -17.8918 & -175.53\\
W & \ding{51} & -12.7685 & -87.3763 & & -13.008 & -87.4066 & \ding{51} & -25.7543 & -174.269\\
Y & & -6.39954 & -86.0638 & & -6.30011 & -85.6655 & & -12.8496 & -172.246\\
Zn & & -1.08594 & -86.1428 & & -1.02165 & -86.0296 & \ding{51} & -2.19499 & -172.505\\
Zr & & -8.46931 & -86.3133 & & -8.42381 & -86.2414 & & -17.0165 & -173.524\\
\end{tabular}
\begin{tablenotes}
\footnotesize
\item $^\dagger$ Dynamically Stabilized (DOI: 10.1103/PhysRevB.95.064101);
\item * The \ding{51} in "Unstable" indicates the phase is mechanically unstable and the "inflection detection" method is applied;
\item ** The "Energy (eV)" is the energy per primitive cell (1 atom per cell for FCC\_A1 and BCC\_A2, 2 atoms per cell for HCP\_A3) at \SI{0}{\kelvin};
\item *** The "Vibrational Entropy" is given per primitive cell, in units of "Boltzmann constant per cell" -- consistent with the convention in ATAT and can be directly used to replace \texttt{svib\_ht}.

\end{tablenotes}
\end{threeparttable}
\end{table}
\end{landscape}

\section{LAMMPS input script for liquid-phase MD calculations using MLIPs}\label{sec:lammpsscript}

\begin{lstlisting}

variable        s equal step
variable        t equal temp
variable        v equal vol
variable        e equal pe
variable        p equal press

units           metal
dimension       3
boundary        p p p
atom_style      atomic

read_data       <STRUCTURE>

pair_style      gnnp <ML-GNNP PATH>
pair_coeff      * *  <MLIP> <VERSION> <ELEMENTS>

timestep        0.001  # 1 fs

# Minimization
reset_timestep  0
thermo          10
thermo_style    custom step temp etotal pe lx ly lz press pxx pyy pzz

min_style       cg
minimize        1e-25 1e-25 5000 10000

reset_timestep  0
velocity        all create <TEMPERATURE> <TEMPERATURE SEED> mom yes rot no

# Equilibration: NPT
fix             1 all npt temp <TEMPERATURE> <TEMPERATURE> $(10.0*dt) iso 0.0 0.0 $(100.0*dt)
thermo          100
thermo_style    custom step temp press pe ke etotal lx ly lz
run             30000  # 30 ps
unfix           1

# Production: NVT
fix             2 all nvt temp <TEMPERATURE> <TEMPERATURE> $(100.0*dt)
fix             Eavg all ave/time 10 3000 30000 etotal file energy ave one
fix_modify      Eavg title1 "" title2 "" title3 ""
thermo          100
thermo_style    custom step temp press pe ke etotal lx ly lz
run             30000  # 30 ps
unfix           2
unfix           Eavg

\end{lstlisting}

\end{document}